 \UseRawInputEncoding
\documentclass[%
 reprint,
superscriptaddress,
floatfix,
 amsmath,amssymb,amsfonts,
pre,
]{revtex4-2}

\usepackage{graphicx}
\usepackage{dcolumn}
\usepackage{bm, bbold}
\usepackage{comment, color}
\usepackage{natbib}
\usepackage{tabularx}
\usepackage{float}
\usepackage{hyperref}

\usepackage{comment}

 \usepackage[ngerman, english]{babel}

\usepackage{float}
\usepackage{lipsum} 

\begin{document}


\def\bea{\begin{eqnarray}}
\def\eea{\end{eqnarray}}
\def\beq{\begin{equation}}
\def\eeq{\end{equation}}
\def\f{\frac}
\def\k{\kappa}
\def\e{\epsilon}
\def\ve{\varepsilon}
\def\be{\beta}
\def\D{\Delta}
\def\h{\theta}
\def\t{\tau}
\def\a{\alpha}

\def\cDa{{\cal D}[X]}
\def\cD{{\cal D}[x]}
\def\cL{{\cal L}}
\def\cLo{{\cal L}_0}
\def\cLa{{\cal L}_1}
\def\rv{{\bf r}}
\def\tv{\hat t}
\def\on{{\omega_{\rm a}}}
\def\od{{\omega_{\rm d}}}
\def\off{{\omega_{\rm off}}}
\def\fv{{\bf{f}}}
\def\fm{\bf{f}_m}
\def\bv{{\bf{v}}}
\def\zh{\hat{z}}
\def\yh{\hat{y}}
\def\xh{\hat{x}}
\def\km{k_{m}}
\def\kk{ {\bf k}}
\def\nh{\hat{n}}

\def\Re{{\rm Re}}
\def\sj{\sum_{j=1}^2}
\def\rk{\rho^{ (k) }}
\def\rek{\rho^{ (1) }}
\def\cek{C^{ (1) }}
\def\rz{\rho^{ (0) }}
\def\rt{\rho^{ (2) }}
\def\rtb{\bar \rho^{ (2) }}
\def\trk{\tilde\rho^{ (k) }}
\def\trek{\tilde\rho^{ (1) }}
\def\trz{\tilde\rho^{ (0) }}
\def\trt{\tilde\rho^{ (2) }}
\def\r{\rho}
\def\tD{\tilde {D}}

\def\bt{{\bf t}}
\def\s{\sigma}
\def\kb{k_B}
\def\bF{\bar{\cal F}}
\def\F{{\cal F}}
\def\la{\langle}
\def\ra{\rangle}
\def\nn{\nonumber}
\def\up{\uparrow}
\def\dn{\downarrow}
\def\S{\Sigma}
\def\dg{\dagger}
\def\d{\delta}
\def\p{\partial}
\def\l{\lambda}
\def\L{\Lambda}
\def\G{\Gamma}
\def\o{\Omega}
\def\w{\omega}
\def\g{\gamma}
\def\E{{\mathcal E}}

\def\O{\Omega}

\def\vv{ {\bf v}}
\def\bn{ {\bf n}}
\def\jv{ {\bf j}}
\def\jr{ {\bf j}_r}
\def\jd{ {\bf j}_d}
\def\jdd{ { j}_d}
\def\noi{\noindent}
\def\a{\alpha}
\def\d{\delta}
\def\p{\partial} 

\def\la{\langle}
\def\ra{\rangle}
\def\e{\epsilon}
\def\n{\eta}
\def\g{\gamma}
\def\break#1{\pagebreak \vspace*{#1}}
\def\hf{\frac{1}{2}}
\def\rcurs{r_{ij}}
\def\rv{ {\mathbf r}}
\def\bv{ {\bf b}}
\def\uv{ {\bf u}}
\def\rv{ {\bf r}}
\def\cf{{\mathcal F}}

\def\fl{\flushleft}    
\def\bea{\begin{eqnarray}}
\def\eea{\end{eqnarray}}
\def\beq{\begin{equation}}
\def\eeq{\end{equation}}
\def\f{\frac}
\def\k{\kappa}
\def\e{\epsilon}
\def\ve{\varepsilon}
\def\be{\beta}
\def\D{\Delta}
\def\h{\theta}
\def\t{\tau}
\def\a{\alpha}

\def\cDa{{\cal D}[X]}
\def\cD{{\cal D}[x]}
\def\cL{{\cal L}}
\def\cLo{{\cal L}_0}
\def\cLa{{\cal L}_1}

\def\Re{{\rm Re}}
\def\sj{\sum_{j=1}^2}
\def\rk{\rho^{ (k) }}
\def\Pe{{\rm Pe}}
\def\rek{\rho^{ (1) }}
\def\cek{C^{ (1) }}
\def\rz{\rho^{ (0) }}
\def\rt{\rho^{ (2) }}
\def\rtb{\bar \rho^{ (2) }}
\def\trk{\tilde\rho^{ (k) }}
\def\trek{\tilde\rho^{ (1) }}
\def\trz{\tilde\rho^{ (0) }}
\def\trt{\tilde\rho^{ (2) }}
\def\r{\rho}
\def\tD{\tilde {D}}

\def\vpl{v_\parallel}
\def\vp{v_\perp}

\def\s{\sigma}
\def\kb{k_B}
\def\bF{\bar{\cal F}}
\def\F{{\cal F}}
\def\la{\langle}
\def\ra{\rangle}
\def\nn{\nonumber}
\def\up{\uparrow}
\def\dn{\downarrow}
\def\S{\Sigma}
\def\dg{\dagger}
\def\d{\delta}
\def\p{\partial}
\def\l{\Pe}
\def\L{\Pe}
\def\G{\Gamma}
\def\o{\Omega}
\def\w{\omega}
\def\g{\gamma}

\def\bv{ {\bf b}}
\def\nv{ \hat{\bm{n}}}
\def\rv{ {\mathbf r}}
\def\vv{ {\mathbf v}}

\def\jv{ {\bf j}}
\def\jr{ {\bf j}_r}
\def\jd{ {\bf j}_d}
\def\jdd{ { j}_d}
\def\noi{\noindent}
\def\a{\alpha}
\def\d{\delta}
\def\p{\partial} 

\def\la{\langle}
\def\ra{\rangle}
\def\e{\epsilon}
\def\n{\eta}
\def\g{\gamma}
\def\break#1{\pagebreak \vspace*{#1}}
\def\hf{\frac{1}{2}}
\def\na{{\eta}_{\rm ac}}
\def\dac{D_{\rm ac}}
\def\n{{\eta}}
\def\gv{\gamma_v}

\def\tbF{\tilde {\bm F}}
\def\tDv{\tilde {D_v}}
\def\tgv{\tilde {\gamma_v}}
\def\tmu{\tilde \mu}

\definecolor{darkmagenta}{rgb}{0.55, 0.0, 0.55}
\definecolor{MyOrange}{RGB}{255, 140, 0}
\newcommand{\alert}[1]{\begingroup\color[rgb]{1,0,0}#1\endgroup}

\title{Finite-Time Orientational Relaxation Restructures Collective Motion in Polar Active Matter}
\author{Rajneesh Kumar}
\email{rajneesh.kumar@iopb.res.in}
\affiliation{Institute of Physics, Sachivalaya Marg, Sainik School, Bhubaneswar 751005, India}
\author{Subhransu Sekhar Mishra }
\email{subhransu.m@iopb.res.in}
\affiliation{Institute of Physics, Sachivalaya Marg, Sainik School, Bhubaneswar 751005, India}
\affiliation{Homi Bhabha National Institutes (HBNI), Training School Complex, Anushakti Nagar, Mumbai, India 400094}
\author{Debasish Chaudhuri}
\email{debc@iopb.res.in}
\affiliation{Institute of Physics, Sachivalaya Marg, Sainik School, Bhubaneswar 751005, India}
\affiliation{Homi Bhabha National Institutes (HBNI), Training School Complex, Anushakti Nagar, Mumbai, India 400094}

\begin{abstract}
We introduce a Langevin formulation of Vicsek-like active particles in which orientations evolve through finite-rate relaxation toward the local mean direction, with alignment strength $J$ and rotational diffusivity $D_r$, thereby combining Vicsek-type local consensus with XY-like orientational dynamics. Using large-scale numerical simulations, we determine the nonequilibrium phase diagram as a function of activity and alignment rate. Increasing the alignment rate drives a sequence of transitions from a homogeneous isotropic state to polar bands, a cross-sea phase of intersecting bands, a homogeneous polar state, and ultimately a micro-clustered regime. The isotropic-to-polar transition is strongly first order, as evidenced by Binder cumulants and bimodal distributions of local polarization and density, indicating coexistence of gas-like and liquid-like regions. Near the onset of collective motion, band size increases with activity but depends non-monotonically on alignment rate. Further increasing the alignment rate drives the system through the cross-sea and homogeneous polar phases before enhanced density fluctuations lead to micro-clustering. Our results demonstrate that finite-time orientational relaxation acts as a control parameter that qualitatively restructures collective behavior in polar active matter.
\end{abstract}


\maketitle

\section{Introduction}

The emergence of collective motion in systems of self-propelled particles is a central problem in nonequilibrium statistical physics. Active matter systems, composed of units that continuously convert energy into motion, display a wide range of collective behaviors across scales, from bacterial suspensions and active colloids to animal flocks and driven granular media~\cite{marchetti2013hydrodynamics, ramaswamy2010mechanics, Palacci2013LivingCO, deseigne2012vibrated, narayan2007long, Bar2020}. A minimal and influential framework to study such phenomena is the Vicsek model~\cite{vicsek1995novel, gregoire2004onset, chate2008collective, vicsek2012collective, Chate2020, ginelli2016physics, chate2022dry}, in which point particles move at constant speed and align their velocities with those of their neighbors in the presence of noise. This simple rule leads to a transition from a disordered isotropic state to a collectively moving polar phase~\cite{vicsek1995novel, gregoire2004onset, chate2008collective}.

Substantial progress has been made in understanding the large-scale behavior of Vicsek-like systems. 
Hydrodynamic theories, supported by numerical simulations, have established true long-range order even in two dimensions~\cite{toner1995long,toner1998flocks,toner2012reanalysis, Toner2022school, chate2008collective}, along with giant number fluctuations~\cite{toner2005hydrodynamics, ginelli2016physics, chate2008collective}, in stark contrast to equilibrium expectations. Complementary kinetic and coarse-grained approaches provide a systematic bridge from microscopic alignment rules to continuum descriptions, clarifying the origin of instabilities and nonlinear structures~\cite{Bertin2006, Bertin2009, farrell2012pattern}. Recent numerical studies and renormalization group calculations suggested a new universality class governing the flocking behavior~\cite{mahault2019quantitative, Jentsch_Lee_2024}.

Recent work has substantially clarified the phase behavior of Vicsek-type models. Numerical studies show that the onset of collective motion is accompanied by phase separation into traveling high-density bands coexisting with a dilute disordered phase, establishing microphase separation and the first-order nature of the transition~\cite{solon2015phase, Solon2015PRE, bhattacherjee2019band}. A crossed-band pattern has been reported deeper in the ordered phase~\cite{bhattacherjee2019band, Chate2020}, and was subsequently identified as a distinct phase, termed the cross-sea phase~\cite{kursten2020dry}. 

A common feature of Vicsek-type models is the assumption of instantaneous alignment: particle orientations are updated to match the local mean direction within a single time step, up to an additive random noise~\cite{vicsek1995novel, gregoire2004onset, chate2008collective, kursten2020dry, Chate2020}. Continuous-time, Langevin-like formulations with local pairwise alignment interactions similar to those of the equilibrium XY model supplemented by self-propulsion, commonly referred to as flying XY models, have also been studied~\cite{farrell2012pattern, chepizko2013collective, liebchen2017collective, Chepizhko2021}.

We consider a Langevin description of active-particle dynamics in which particle orientations relax toward the mean heading of their local neighborhood. As in the Vicsek model~\cite{vicsek1995novel}, alignment is neighborhood-averaged rather than pairwise, distinguishing it from flying XY models~\cite{farrell2012pattern, chepizko2013collective, liebchen2017collective, Chepizhko2021}. However, unlike the instantaneous reorientation of the Vicsek model, heading directions evolve continuously at a finite alignment rate, introducing an orientational relaxation timescale. This framework combines Vicsek-type alignment with continuous-time Langevin dynamics while providing independent control over activity and alignment, thereby enabling a systematic investigation of their role in collective motion.

The resulting phase diagram captures the full range of structural states in active polar systems, including homogeneous isotropic, polar banded, cross-sea, homogeneous polar, and polar micro-clustered phases. The isotropic-polar transition is strongly first-order, marked by coexistence of low-density gas-like and high-density liquid-like regions that persist even at modest system sizes.
Beyond the phase diagram, we characterize the structural and fluctuation properties of the ordered states, revealing a nonmonotonic dependence of density fluctuations on alignment rate and systematic variations in band width with activity and alignment. Overall, our results establish that a minimal Langevin framework is sufficient to capture the rich phenomenology of Vicsek-like systems, linking microscopic alignment dynamics to emergent collective motion and pattern formation.

The remainder of the paper is organized as follows. In Sec.~\ref{sec_model}, we introduce the model and simulation methods. Section~\ref{sec:results} presents the main results, including the phase diagram, the nature of the transition, the properties of the orientationally ordered phase, and the structural characterization of the system. Finally, in Sec .~\ref{sec:conclusion}, we conclude with an outlook.

\begin{figure*}[t]
\centering
  \includegraphics[width=16cm]{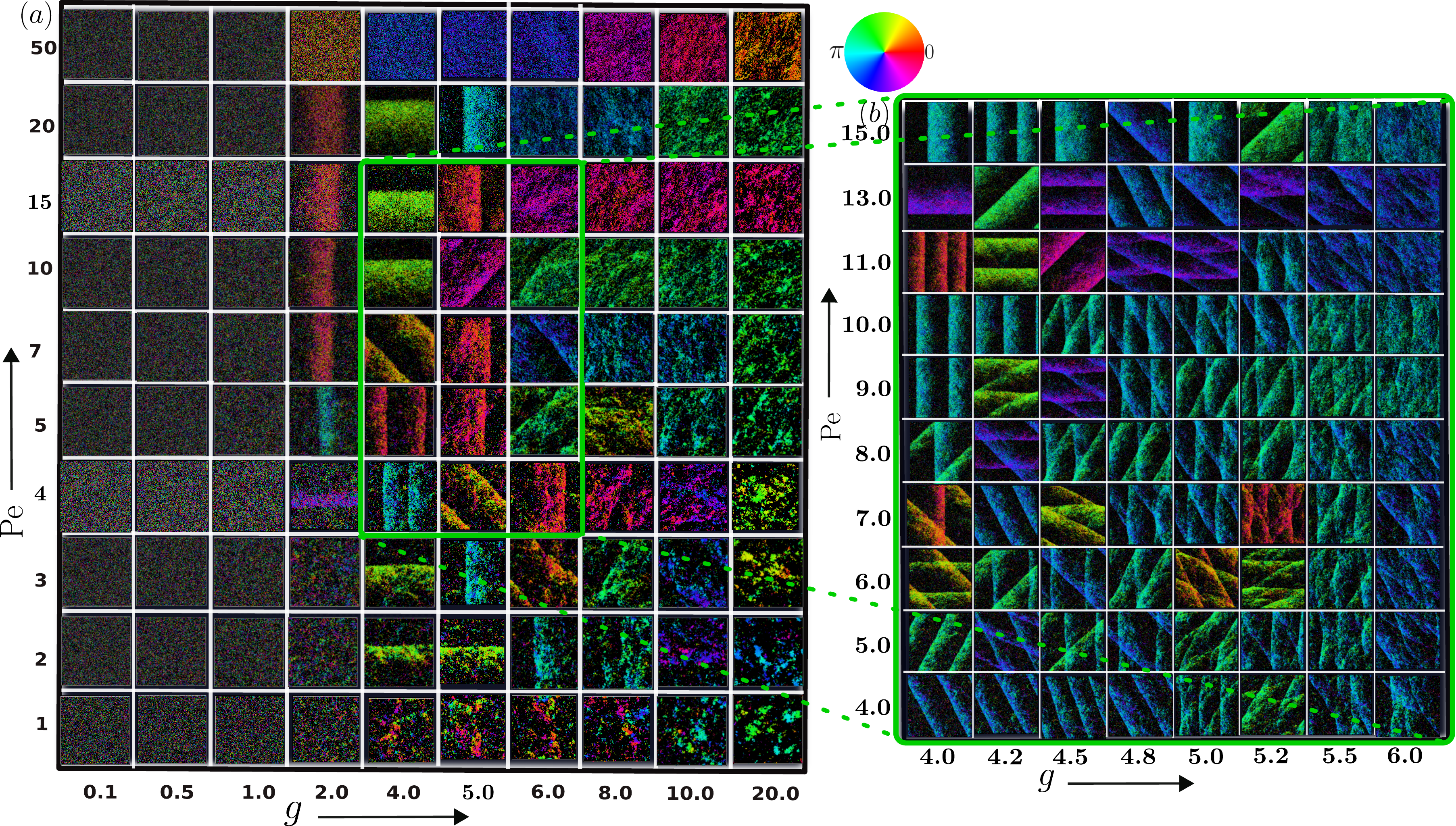} 
 \caption{
 (a) Configurational phase diagram in the $(\Pe, g)$ plane for a system of $N = 16{,}000$ particles at fixed number density $\rho = 1.0$. Particle colors represent the heading direction $\mathbf{u}_i = (\cos \theta_i, \sin \theta_i)$, with $\theta_i \in [0, 2\pi]$. Variation of $\Pe$ and $g$ gives rise to pattern formation associated with an order-disorder transition, including homogeneous isotropic states, polar bands, polar micro-clusters, and homogeneous polar phases.
(b) Configurational phase diagram in the $(\Pe, g)$ plane for a system with $N=64{,}000$ particles, corresponding to the parameter region highlighted by the green box in (a). The enlarged system size reveals an additional cross-sea phase that is absent in smaller systems.
}\label{fig_PD_config}
\end{figure*}

\section{Model}\label{sec_model} 

A common feature of Vicsek-type models is the assumption of instantaneous alignment, whereby particle orientations are updated to the local mean direction within a single time step. In many active systems, however, alignment is expected to proceed over a finite timescale. To account for this effect, we introduce a continuous-time Vicsek-like model in which alignment arises through local relaxation dynamics. Interpreting the stochastic evolution in the It{\^o} sense, the orientation $\theta_i$ of particle $i$ obeys
\begin{equation}
d\theta_i = - J \sin(\theta_i-\psi_i)\,dt + \sqrt{2D_r}\,dB_i,
\end{equation}
where $J=\tau_{\rm align}^{-1}$ sets the alignment rate and
\begin{equation}
\psi_i=\tan^{-1}\left(\frac{\sum_{j=1}^{n_i}\sin\theta_j}{\sum_{j=1}^{n_i}\cos\theta_j}\right)
\end{equation}
is the instantaneous mean orientation of the $n_i$ particles within an interaction radius $r_c$, including particle $i$. Here $D_r$ denotes the rotational diffusivity and $dB_i$ is a Wiener increment satisfying $\langle dB_i(t)\rangle=0$ and $\langle dB_i(t)dB_j(t')\rangle=\delta_{ij}\delta(t-t')\, dt$. The inverse diffusivity, $D_r^{-1}$, therefore sets the orientational persistence time of an isolated particle. Particle positions evolve according to
\begin{equation}
d\mathbf{r}_i=v_0\,\mathbf{u}_i\,dt,
\end{equation}
where $v_0$ is the self-propulsion speed and $\mathbf{u}_i=(\cos\theta_i,\sin\theta_i)$. In the limit $J\rightarrow\infty$ ($\tau_{\rm align} \to 0$), orientational relaxation becomes effectively instantaneous, recovering the continuous-time counterpart of the Vicsek model (Appendix~\ref{app:vicsek_limit}).

The finite alignment rate introduces a relaxation timescale that enables systematic control of delayed alignment. We characterize the system in terms of two dimensionless parameters: the effective alignment rate $g=J/D_r$ and the P{\'e}clet number $\mathrm{Pe} = v_0/(D_r r_c)$, which compares active advective transport to rotational diffusion.

Using $\tau_p = 1/D_r$ and $r_c$ as units of time and length, respectively, and rescaling $\mathbf{r}_i \to \mathbf{r}_i/r_c$ and $t \to D_r t$, the equations of motion become
\begin{equation}
d \mathbf{r}_i = \mathrm{Pe}\, \mathbf{u}_i\, dt ,
\label{eq_r}
\end{equation}
\begin{equation}
d\theta_i = - g \sin(\theta_i - \psi_i)\, dt + \sqrt{2}\, dB_i .
\label{eq_h}
\end{equation}

We integrate these equations using the Euler-Maruyama scheme at fixed number density $\rho = N/L^2$ for $N$ particles in a square domain of size $L$ with periodic boundary conditions. The updates use forward Euler discretization for both position and orientation. Similar to this, the original Vicsek model employs a discrete-time map with updates akin to a forward Euler scheme~\cite{vicsek1995novel}, whereas several later variants combine forward and backward Euler updates~\cite{chate2008collective, kursten2020dry, Chate2020}.

\begin{figure}[t!]
\centering
    \includegraphics[width=8.6cm]{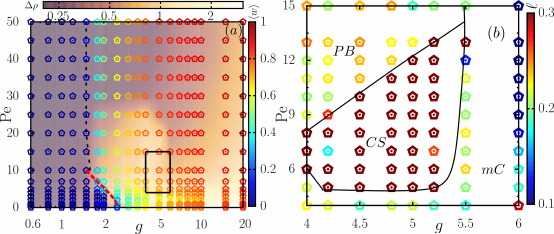}
 \caption{  Phase diagram in the $(\Pe, g)$ plane,  where the dimensionless activity is defined as $\Pe = v_0/(r_c D_r)$ and the effective alignment rate as $g = J/D_r$.
(a)~Points are colored by the polar order parameter $\la  w  \ra$, while the background heat map represents density fluctuations $\D \r$. The black dashed line marks the order-disorder phase boundary, determined from the maxima of fluctuations in the polar order parameter. The red dashed line represents $\Pe_\ast = ({\cal B}/g_\ast)(g_\ast^0-g_\ast)^\hf$, with ${\cal B}=11$ and $g_\ast^0=2.5$, corresponding to Eq.~\eqref{eq:v0_boundary}.
(b)~Enlarged view of the region identified by the black rectangle in (a), showing the cross-sea phase, polar banded phase, and microcluster phase. Points are colored according to the structural order parameter $\langle \ell \rangle$, as defined in Sec.\ref{sec_VL}.
 }\label{fig_PD}
\end{figure}

\begin{figure}[t]
\centering 
  \includegraphics[width=8.5cm]{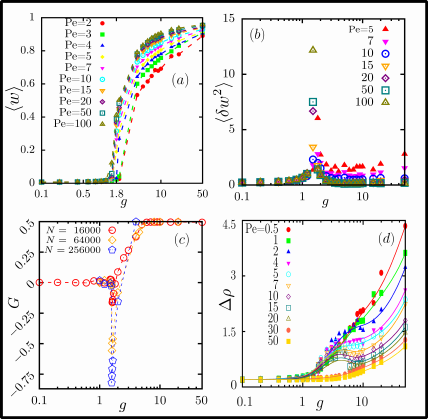}
 \caption{
 Steady-state properties as a function of the dimensionless alignment rate $g$ for different activities $Pe$. Results are shown for $N=16{,}000$ particles at number density $\rho=1.0$.
(a) Polar order parameter $\langle  w  \rangle$ versus $g$.
(b) Variance of the polar order, $\langle \delta  w ^2 \rangle$, versus $g$, showing peaks near the order-disorder transitions.
(c) Binder cumulant $G$ versus $g$ for $Pe=10$. A negative dip near $g^\ast$ signals a first-order transition; colors indicate different system sizes.
(d) Root-mean-square density fluctuations $\D \rho$ versus $g$.
}\label{fig_PvsJ}
\end{figure}

\section{Results}\label{sec:results}

\subsection{Phase behaviour and nature of transition}\label{sec:PD}

\textit{Phase diagram.} We first characterize the steady states of self-propelled aligning particles in a square domain with periodic boundary conditions. Figure~\ref{fig_PD_config} shows the configurational phase diagram in the $(\Pe,g)$ plane, where $\Pe = v_0/(r_c D_r)$ is the dimensionless activity and $g = J/D_r$ denotes the effective alignment rate controlling the transition (see Appendix~\ref{app_alignment}). The orientational order--disorder transition is accompanied by a wide range of emergent patterns, giving rise to multiple structural phases within the polar-ordered regime.
For a system with $N=16{,}000$ [Fig.~\ref{fig_PD_config}(a)], four distinct dynamical regimes are identified.

At low $g$ and low $\Pe$, the system remains in a homogeneous isotropic (HI) phase with negligible global alignment. At sufficiently large $\Pe$ and moderate $g$ (e.g., $g=5$, $\Pe=50$), a homogeneous polar (HP) phase emerges, characterized by global alignment and weak density inhomogeneities. Increasing $g$ in this regime enhances density fluctuations despite the persistence of global order. At large $g$, the system enters a microclustered phase (mC), characterized by flocks of limited spatial extent.

In a broad region of parameter space, the transition from the disordered to the ordered state is mediated by a traveling polar banded (PB) phase, consisting of dense, coherently moving bands coexisting with a dilute background. The morphology of this phase depends sensitively on both control parameters: increasing $\Pe$ broadens the bands and can eventually make them span the system -- a finite-size effect confirmed in larger systems -- whereas increasing $g$ increases the number of bands, which subsequently merge into wider structures. This coarsening process exhibits a pronounced dependence on system size.

The phase behavior varies qualitatively with activity. For small $\Pe$, the banded phase is absent, and the system transitions directly from the HI phase to an mC state as $g$ increases. In this regime, clusters initially move in different directions and progressively align at larger $g$. In contrast, at very large $\Pe$, the system exhibits a direct transition from the HI phase to the HP phase, bypassing band formation.

For a larger system size, $N=64{,}000$, an additional regime -- the cross-sea (CS) phase -- stabilizes over an intermediate range of $(\Pe,g)$ [Fig.~\ref{fig_PD_config}(b)]. This observation is consistent with Refs .~\cite {kursten2020dry,bhattacherjee2019band}. In this regime, increasing $g$ first leads to multiple coexisting bands, which at higher $g$ interact strongly and organize into intersecting, grid-like patterns characteristic of the CS phase~\cite{Chate2020}. A detailed structural and quantitative characterization of this state is presented later in Sec.~\ref{sec_VL}.

It is evident from the above discussion that the phase transition entails the emergence of orientational order in the heading directions, along with phase separation in the density field. A complementary representation of the phase diagram is shown in Fig.~\ref{fig_PD}(a), where the color scale of the points denotes the scalar polarization order parameter  
$\langle  w  \rangle = \la |\, \sum_{i=1}^N {\bf u}_i /N \,|\ra$ with $\la \dots \ra$ denoting a steady-state time average~\cite{vicsek1995novel, gregoire2004onset, chate2008collective}. The underlying heat map in the same plot encodes density fluctuations, quantified by $\D \rho = [\langle \rho^2 \rangle - \langle \rho \rangle^2]^{1/2}$, thereby capturing spatial inhomogeneities. These fluctuations remain small at the onset of polar order, marked by the order-disorder phase boundary (dashed black line) obtained from the loci of the maxima in the order-parameter variance. In the weak-activity regime, the boundary follows the scaling $\Pe_\ast = ({\cal B}/g_\ast)(g_\ast^0-g_\ast)^{1/2}$ predicted by Eq.~\eqref{eq:v0_boundary}, while crossing over to an activity-independent form at large $\Pe$. At finite activity, orientational noise in the heading direction generates positional fluctuations that vanish in the equilibrium limit of vanishing $\Pe$. To account for finite diffusion in this limit, we include an additional translational noise.

Pronounced density fluctuations arise in two regimes: (i) at intermediate $(\Pe,g)$, encompassing the cross-sea (CS) phase, and (ii) at large $g$, corresponding to the microclustered regime. In contrast, large $\Pe$ generally suppresses density variations, as quantified below.
Figure~\ref{fig_PD}(b) delineates the boundaries of the polar banded (PB), microcluster (mC), and cross-sea (CS) phases for a system of size $N=64{,}000$. The values of the auxiliary order parameter $\ell$, introduced later to characterize the CS phase, are indicated by color-coded markers.

\textit{Order parameter and phase transition.} The dependence of $\langle  w  \rangle$ on $g$ for different $\Pe$ [Fig.~\ref{fig_PvsJ}(a)] reveals a transition from a disordered state ($\la  w  \ra = 0$) to an ordered state ($\la  w  \ra \neq 0$) beyond a threshold $g^\ast$, which decreases with increasing $\Pe$. The corresponding variance, $\la \delta  w ^2 \ra= \langle  w ^2 \rangle - \langle  w  \rangle^2$, exhibits a pronounced peak at the transition [Fig.~\ref{fig_PvsJ}(b)], with larger fluctuations observed at higher $\Pe$. The phase boundary, shown as the dashed black line in Fig.~\ref{fig_PD}(a), is obtained from the loci of these variance maxima.

To investigate the phase transition further, we compute the Binder cumulant of the polar order, $G = 1 - \langle  w ^4 \rangle / 2 \langle  w ^2 \rangle^2$~\cite{Binder1997}, across the ordering transition for various system sizes (Fig.~\ref{fig_PvsJ}(d)\,). This quantity is defined such that for a Gaussian distribution with $\langle  w  \rangle = 0$, the cumulant vanishes, and in the fully ordered phase, it approaches $1/2$. For a first-order transition, $G$ develops a negative minimum near the transition point that sharpens with increasing system size, signaling coexistence of the ordered and disordered phases~\cite{Vollmayr1993}. As shown in Fig.~\ref{fig_PvsJ}(c), the minimum becomes progressively deeper for larger systems, providing clear evidence of a first-order transition at $g^\ast$.

\textit{Density fluctuations.} Additional insight is obtained from the behavior of $\D \rho =  [\langle \rho^2 \rangle - \langle \rho \rangle^2]^\hf$ as a function of $g$ [Fig.~\ref{fig_PvsJ}(d)]. For small $\Pe$, density fluctuations remain weak at low $g$ and increase monotonically, indicating gradual cluster formation. At intermediate $\Pe$, fluctuations grow sharply near the transition and peak in the PB and CS phases, reflecting strong density segregation. Upon further increase in $g$, band and cross-sea structures broaden and merge, leading to a transient reduction in fluctuations, followed by a renewed increase as microclusters emerge and grow. The consistently large value of $\D \rho$ deep inside the ordered phase is consistent with the expectation of giant number fluctuations, which we discuss below. For large $\Pe$, the PB phase is absent, and fluctuations increase smoothly with $g$, consistent with a direct crossover to a homogeneous polar state that progressively develops density inhomogeneities.

\subsection{Mean field analysis}
{\it Transition point.} Starting from the Langevin dynamics in Eq.~\eqref{eq_h}, the probability distribution of particle orientations obeys the Fokker--Planck equation~\cite{Chepizhko2021, Sinha2023}
\begin{equation}
\partial_t p(\theta,t)
=\partial_\theta \!\left[J w \sin(\theta-\psi)\, p(\theta,t) \right]
+
D_r \partial_{\theta}^2 p(\theta,t).
\label{eq:fp}
\end{equation}
where
$
w =\left| \int_0^{2\pi} d\theta \, p(\theta)e^{i\theta} \right|
$
is the polar order parameter with mean orientation $\psi$, $J$ is the alignment strength, and $D_r$ is the rotational diffusivity. The homogeneous steady-state solution of Eq.~\eqref{eq:fp} is the von Mises distribution,
\begin{equation}
p(\theta)
=
\frac{1}{2\pi I_0(\kappa)}
\exp\!\left[\kappa \cos(\theta-\psi)\right],
\label{eq:vm}
\end{equation}
with $\kappa = g  w $, where $g=J/D_r$ and $I_n$ denotes the modified Bessel function of the first kind of order $n$. 
Substituting Eq.~\eqref{eq:vm} into the definition of $w$ yields the self-consistency condition
$  w =I_1(\kappa)/I_0(\kappa)$. Expanding for small $\kappa$,
$ I_1(\kappa)/I_0(\kappa) = \kappa/2 - \kappa^3/16 + O(\kappa^5),$
and using $\kappa=g  w $, it reduces to  
$ (g-g_c)  w -\left(g/g_c\right)^3  w ^3=0, $
with $g_c=2$. Hence, $ w =0$ for $g\le g_c$, while for $g>g_c$,
\bea
 w  = \left(g_c/g\right)^{3/2} (g-g_c)^{1/2}.
\eea

The homogeneous mean-field solution therefore predicts a continuous transition at $g_c=2$, independent of activity and density, with critical exponent $\beta=1/2$. Activity-induced density fluctuations, however, enhance ordering, drive the transition first-order, and shift the transition to lower alignment rates, $g_\ast<g_c$~\cite{Martin2024flucind, Sinha2023}. 

{\it Active suppression of the transition point.}
To rationalize the activity dependence of the transition, we consider a one-dimensional form of the Toner--Tu equations~\cite{caussin2014emergent,Solon2015a}. The local density $\rho$ and polarization $w$ obey
\begin{align}
\partial_t \rho + v_0 \partial_x w &= 0,
\nonumber\\
\partial_t w + \xi w \partial_x w
&=
-\frac{\partial f}{\partial w}
-\lambda \partial_x \rho
+\nu \partial_{xx} w ,
\label{eq:TT1D}
\end{align}
with local free-energy density~\cite{farrell2012pattern}
\begin{equation}
f(w)
=
-\frac{1}{2}\left(\frac{g\rho}{2}-1\right)w^2
+\frac{a_4}{4}w^4 .
\label{eq:fw0}
\end{equation}

Near the onset of order, the system develops traveling bands propagating through a disordered background. Assuming traveling-wave solutions $\rho(z)$ and $w(z)$ with $z=x-ct$, the continuity equation yields~\cite{Solon2015a}
\begin{equation}
\rho
=
\bar{\rho}
+
\frac{v_0}{c}w .
\label{eq:rho_band}
\end{equation}
Substituting Eq.~\eqref{eq:rho_band} into Eq.~\eqref{eq:fw0} gives the effective free-energy density
\begin{equation}
f(w)
=
-\frac{1}{2}
\left(
\frac{g\bar{\rho}}{2}-1
\right)w^2
-
\frac{gv_0}{4c}w^3
+
\frac{a_4}{4}w^4 .
\label{eq:fw_eff}
\end{equation}
The nonequilibrium advective coupling therefore generates a cubic term proportional to $v_0$, which drives the transition first-order~\cite{Sinha2024}.
Using the condition of coexistence of disordered gas ($w_g=0$) and the ordered liquid ($w_l\neq0$), $\frac{\partial f}{\partial w}=0$, and  $f(w_l)=f(w_g=0)=0$, we obtain 
\begin{equation}
w_l = \frac{v_0 g_\ast}{2a_4 c}, \qquad   \bar{\rho} g_\ast = 2 - \frac{v_0^2 g_\ast^2}{4a_4 c^2}.
\label{eq:gstar}
\end{equation}
In the weak-activity limit,
$
\bar{\rho} g_\ast \simeq 2 - \frac{v_0^2}{a_4 c^2 \bar{\rho}},
$
demonstrating that activity shifts the transition below its equilibrium value $\bar{\rho}g_c=2$, recovered in the limit $v_0=0$. 
The liquid--gas coexistence is purely activity-induced: a disordered gas with $w_g=0$ and $\rho_g=\bar{\rho}$ coexists with an ordered liquid characterized by $w_l=v_0 g_\ast/(2a_4c)$ and $\rho_l= \rho_g + (v_0/c)w_l$. 
The phase boundary $(g_\ast, v_\ast)$ obeys Eq.~\eqref{eq:gstar} giving 
\begin{equation}
v_\ast = \frac{2c\sqrt{a_4}}{g_\ast} \sqrt{ 2-\bar{\rho}g_\ast}.
\label{eq:v0_boundary}
\end{equation}
Expressed in terms of the dimensionless activity $\Pe$, this relation yields an estimate for the phase boundary at $\Pe_\ast$ and $g_\ast$ given by $\Pe_\ast = (\mathcal{B}/g_\ast)(g_\ast^0 - g_\ast)^{1/2}$. This expression agrees well with the low-activity behavior and captures the activity-induced suppression of the transition (Fig.~\ref{fig_PD}). At larger activity, however, the phase boundary becomes nearly independent of $\Pe$, indicating the need for a more refined theoretical description.

\subsection{Nature of the ordered state}\label{sec:polar}

\begin{figure}[]
\includegraphics[width=8cm]{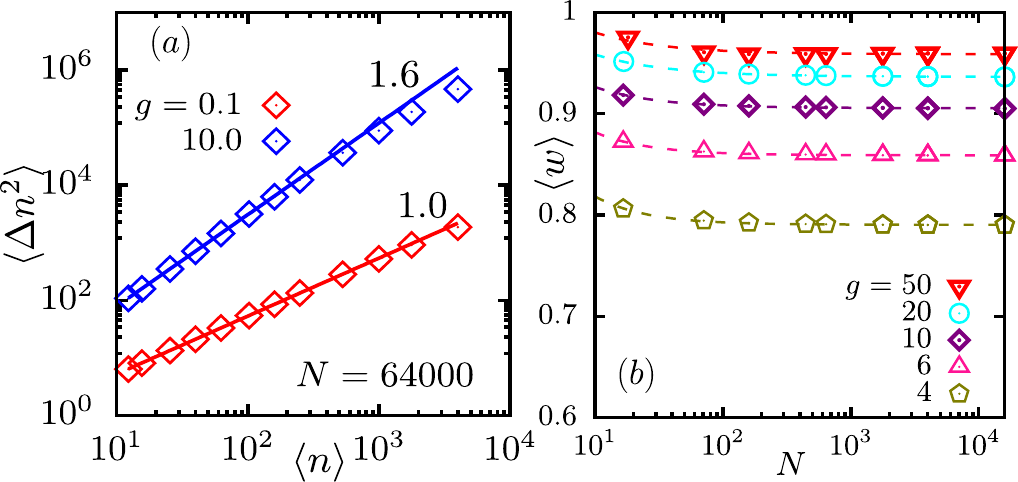}
\caption{
(a) Scaling of number fluctuations, $\langle \Delta n^2\rangle \sim \langle n\rangle^\alpha$, for $\Pe=10$ in the disordered phase ($g=0.1$) and homogeneous polar phase ($g=10$). The lines show $\alpha=1$ for $g=0.1$ and $\alpha=1.6$ for $g=10$, demonstrating giant number fluctuations in the ordered phase. Results are for $N=6.4\times10^4$ and $\rho=1$. (b) Polar order parameter $\langle w \rangle$ versus system size $N$ for different alignment rates $g$ at $\Pe=50$. Dashed lines are fits to $\la  w  \ra =  w _\infty+c/N^\nu$. The fitted $( w _\infty,\nu)$ values for $g=4,6,10,20,50$ are $(0.79,1.0)$, $(0.86,0.99)$, $(0.90,0.89)$, $(0.94,0.76)$, and $(0.96,0.80)$, respectively.
}\label{fig_GNF}
\end{figure}

\textit{Giant number fluctuation.}  We quantify number fluctuations by measuring the mean-squared fluctuations, $\la \Delta n^2 \ra=\langle n^2\rangle-\langle n\rangle^2$, for subregions with mean occupancy $\langle n\rangle$, and examine their scaling with $\la n \ra$~(Fig.\ref{fig_GNF}(a)). At $\Pe=10$, we find $\la \Delta n^2 \ra\sim \la n\ra^\alpha$, with $\alpha$ depending on the phase: in the disordered phase ($g=0.1$), $\alpha = 1$, consistent with normal (Poissonian) fluctuations, whereas in the homogeneous polar phase ($g=10.0$) we obtain $\alpha = 1.6$. The latter clearly indicates the presence of giant number fluctuations (GNF), and the exponent $\alpha = 1.6$ is consistent with earlier results~\cite{chate2008collective} and a theoretical argument based on the Toner-Tu theory~\cite{ginelli2016physics, toner1998flocks}.

\textit{Long-range order.}  The existence of long-range order in two-dimensional polar flocks has been established by the Toner-Tu theory and confirmed in previous numerical studies~\cite{chate2022dry, Toner2022school}. Although the Mermin--Wagner--Hohenberg theorem forbids spontaneous symmetry breaking in equilibrium two-dimensional systems, active matter circumvents this restriction through its intrinsically nonequilibrium dynamics~\cite{Toner2022school}. To examine the stability of the homogeneous polar state in our system, we study the finite-size dependence of the global order parameter $\langle w \rangle$, shown in Fig.~\ref{fig_GNF}(b). For all alignment rates in the ordered regime, $\langle w \rangle$ approaches a finite asymptotic value following $\la  w  \ra -  w _\infty \sim N^{-\nu}$, indicating the persistence of long-range polar order $ w _\infty$ in the thermodynamic limit. Similar behavior was reported previously~\cite{Chate2020}; however, unlike the constant exponent found there, our results reveal a continuously varying exponent $\nu(g)$ that decreases with increasing $g$ (Fig.~\ref{fig_GNF}(b)), indicating slower convergence toward the asymptotic macroscopic value deeper in the ordered phase. The decay of $\langle w \rangle - w_\infty$ follows an alignment-rate-dependent algebraic form similar to that characterizing quasi-long-range nematic order in active nematics~\cite{ramaswamy2003active, toner2005hydrodynamics}.
 Furthermore, the asymptotic polar order parameter $w _\infty$ in our system increases systematically with alignment rate, from $ w _\infty\simeq0.79$ at $g=4$ to $ w _\infty\simeq0.96$ at $g=50$, reflecting the enhanced coherence of collective motion at stronger alignment.

Overall, the interplay between activity and alignment governs both the phase structure and the first-order character of the ordering transition.

\begin{figure}[t]
\centering
\includegraphics[width=8.5cm]{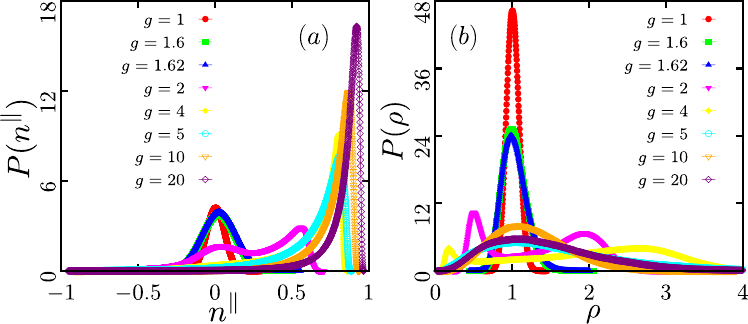}
 \caption{For a system of size $N = 16{,}000$ at $\Pe = 10$, the local probability distributions (a) $P(n^\parallel)$ and (b) $P(\rho)$ are shown for various values of the dimensionless alignment rate $g$. In both cases, the distributions exhibit clear bimodality near transition, indicating phase coexistence.
    }\label{fig_LocalProbPRhope10}
\end{figure}

\subsection{Local ordering analysis}\label{local_ordering}

To characterize the local organization, we analyze two local observables computed from steady-state configurations. First, we consider the projection of the tagged-particle heading direction onto the instantaneous local polarization,
$n_i^\parallel = \hat{\mathbf n}_i \cdot \mathbf w_i ,$
where
$\mathbf w_i = \frac{1}{\nu_i}\sum_{j\in \partial_i}\hat{\mathbf n}_j$
is the local polarization within a neighborhood $\partial_i$ of radius $r_0=10r_c$, and $\nu_i$ is the corresponding number of neighbors. Second, we define the local density as
$\rho_i = {\nu_i}/ ({\pi r_0^2}).$
The corresponding probability distributions, evaluated over all particles in steady state, are denoted by $P(n^\parallel)$ and $P(\rho)$, respectively.

As shown in Fig.~\ref{fig_LocalProbPRhope10} for $\Pe=10$ and $N=16000$, both $P(n^\parallel)$ and $P(\rho)$ evolve from unimodal to bimodal forms with increasing $g$, indicating the emergence of local ordering accompanied by phase separation. The transition is therefore characterized by coexistence between a polar fluid and an isotropic gas.

At low $g$, the system remains spatially homogeneous. Accordingly, $P(n^\parallel)$ is sharply peaked near zero, reflecting the absence of local orientational order, while $P(\rho)$ is centered around the mean density $\rho=1$ with relatively small fluctuations.

As $g$ increases, a second peak develops in $P(n^\parallel)$ at larger values of $n^\parallel$, signaling the appearance of locally aligned domains embedded within a disordered background. The progressive growth and eventual dominance of this peak mark the onset of global polar order. Simultaneously, $P(\rho)$ acquires a bimodal structure: the low-density peak shifts toward smaller densities, whereas the high-density peak moves to larger values, corresponding respectively to dilute regions and dense traveling bands.

At sufficiently large $g$, $P(n^\parallel)$ becomes unimodal again, now with a sharp peak at large $n^\parallel$, consistent with a globally aligned state. In contrast, $P(\rho)$ remains broad with a pronounced high-density tail, indicating the persistence of density inhomogeneities in the form of microclusters whose characteristic size grows with increasing $g$.

\begin{figure}[t]
\centering
 \includegraphics[width=4cm]{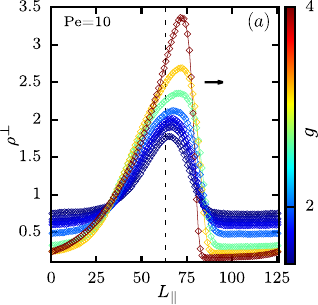}  
  \includegraphics[width=4cm]{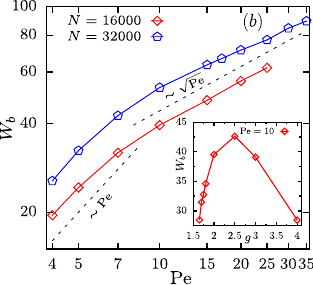} 
\caption{
Time-averaged band profiles and widths. (a) Local density along the polarization direction for different $g$ at $N=16{,}000$, recentered on the largest cluster; arrows indicate motion and dashed lines mark the box midpoint.  (b) Band width $W_b$ versus activity $\Pe$ for $N=16{,}000$ and $32{,}000$ at $g=2$, scaling linearly at low $\Pe$ and as $\sim\sqrt{\Pe}$ at high $\Pe$; inset shows $W_b$ versus $g$ at $\Pe=10$.
}
\label{fig_DenProfileg}
\end{figure}

\subsection{Width of the bands}\label{BandWidth}
The time-averaged density profiles along the direction of band propagation are used to quantify the effects of activity $\Pe$ and alignment rate $g$ on band morphology. The analysis is performed for two system sizes, $N=16{,}000$ and $N=32{,}000$, with profiles recentered by aligning the center of mass of the largest cluster with the center of the simulation box. The local density is obtained by binning along the propagation direction, with the box midpoint indicated for reference. Representative profiles at $N=16{,}000$ for different values of $g$ are shown in Fig.~\ref{fig_DenProfileg}(a). As $g$ increases, the peak density rises, reflecting denser and more sharply localized bands. 

We next examine the dependence of the band width on activity. As shown in Fig.~\ref{fig_DenProfileg}(b) for fixed $g=2$, the density profiles broaden systematically with increasing $\Pe$, and this trend is consistent across both system sizes considered, $N=16{,}000$ and $32{,}000$. To quantify the broadening, we extract the band width $W_b$ from the full width at half maximum of the density profiles. The resulting behavior reveals two distinct scaling regimes: at low activity, $W_b$ increases approximately linearly with $\Pe$, while at larger activity it crosses over to a slower growth consistent with $W_b\sim\sqrt{\Pe}$. Importantly, this scaling behavior is essentially independent of system size, indicating that the observed broadening is an intrinsic property of the banded state rather than a finite-size effect.

The band width, however, displays a nonmonotonic dependence on $g$: it increases up to $g\approx 2.5$ and decreases for larger $g$ (Fig.~\ref{fig_DenProfileg}(b)). For strong alignment, $g>4$, the coherent band destabilizes and fragments into multiple microclusters.

\begin{figure}[t]
\centering
 \includegraphics[width=4.75cm]{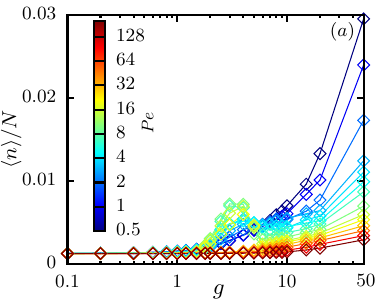} 
 \includegraphics[width=3.8cm]{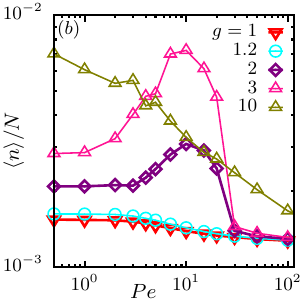} 
  \caption{Fraction of average cluster size $\langle n \rangle /N$ as a function of alignment rate $g$ for various $\Pe$ values, with system size $N=16000$ and density $\rho=1$. Colors denote different $\Pe$. (b) $\langle n \rangle /N$ as a function of $\Pe$ for fixed $g$, spanning disordered, banded, and microcluster regimes. 
  }\label{fig_AveN}
\end{figure}

\subsection{Cluster size analysis}\label{cluster}
{\it Mean cluster size.} In Fig.~\ref{fig_AveN}(a), we present the average cluster size, $\langle n \rangle/N$ -- the fraction of particles belonging to a typical cluster -- as a function of the alignment rate $g$ for different activity values $\Pe$. The results reveal an overall increase in cluster size with increasing $g$, consistent with the progression from a disordered state to banded and subsequently microcluster-dominated phases. In particular, for intermediate values of $g$, corresponding to the banded phase stabilized by finite $\Pe$, the mean cluster size attains a maximum. As $g$ is increased further, $\langle n \rangle$ decreases as the system transitions into the microcluster phase, where particles are distributed more homogeneously among many smaller clusters. In the strongly aligning regime, however, the mean flock size $\langle n \rangle$ increases once again, reflecting an increase in the size of the microclusters.

Figure~\ref{fig_AveN}(b) shows that the mean cluster size generally decreases with increasing activity. Nevertheless, this dependence is non-monotonic, with a pronounced maximum emerging in the parameter regime corresponding to the PB and CS phases. In the large-alignment regime, $g=10$, however, $\langle n \rangle$ decreases monotonically with $\Pe$, indicating a progressive reduction in cluster size as the system evolves toward a homogeneous polar fluid. This behavior is facilitated by stronger active fluctuations within the polar-ordered phase.

\begin{figure*}[t]
\centering
 \includegraphics[width=16cm]{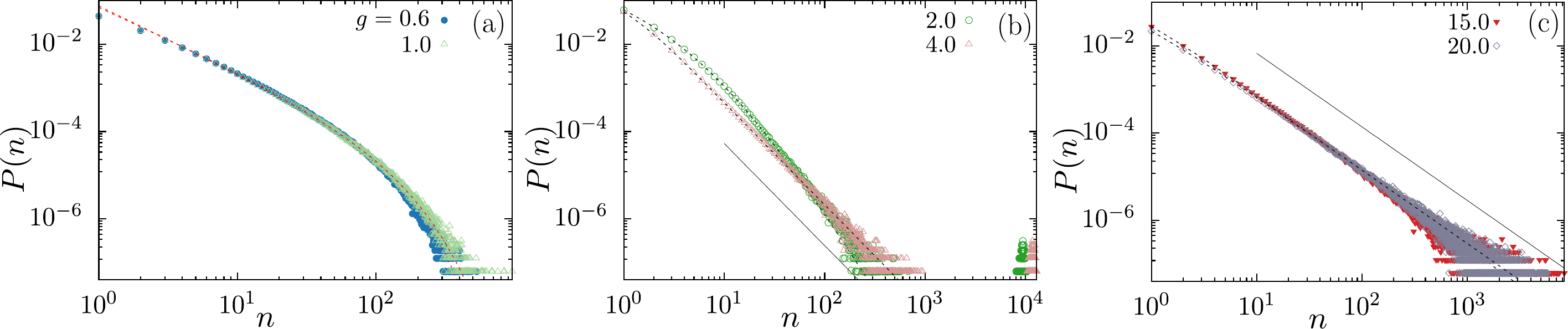} 
 \caption{ Cluster-size distribution $P(n)$ for various $g$ at $\mathrm{Pe}=10$. The simulation data are shown as points, while the dashed lines represent the corresponding fitted functions.
(a) Disordered phase: The distributions are well described by a form $e^{-n/n^\ast} n^{-\beta}$~(Eq.\eqref{eq_exp}), with characteristic sizes $n^\ast=68$ and $85$, and exponents $\be=1.46$ and $1.53$  for $g=0.6$ and $1$, respectively. 
(b),(c) Polar phase: The distributions are described by $n^{-\alpha}\bigl[1+(n/n_c)^{\gamma}\bigr]^{(\alpha-\beta)/\gamma}$~(Eq.\eqref{eq_power}).  The corresponding asymptotic exponents are $\beta\, (g)=3.76 \,(2),\,  2.34\, (4),\, 1.76 \,(15),\, 1.68 \,(20)$. The black solid lines indicate the scaling $n^{-\beta}$, with $\beta=2.34$ for $g=4$ in panel (b) and $\beta=1.68$ for $g=20$ in panel (c).
  }\label{fig_Cn}
\end{figure*}

{\it Cluster size distribution.} The cluster-size distribution $P(n)$ at $\Pe=10$ is shown in Fig.~\ref{fig_Cn} for different $g$. Here $P(n)$ is normalized such that $\sum_{n=1}^{N} n P(n)=N$. In the disordered phase (Fig.~\ref{fig_Cn}(a)), the distribution is well described by
\begin{equation}
g(n)=\mathcal{A}\, n^{-\be}\exp(-n/n^\ast),
\label{eq_exp}
\end{equation}
indicating an exponential cutoff at a characteristic size $n^\ast$, modulated by a power-law prefactor with exponent $\be$. This form is consistent with kinetic-theory predictions~\cite{Peruani2013} and experimental observations in bacterial colonies~\cite{Zhang2010}. Within the homogeneous phase,  as $g$ increases, $n^\ast$ grows and the distribution approaches a power-law form while the exponent remains close to $\be\simeq 3/2$ (Fig.~\ref{fig_Cn}(a)).

Upon entering the polar phase, $P(n)$ crosses over to a scale-free form that is accurately captured by the double power-law expression
\begin{equation}
f(n)=A n^{-\alpha}\left[1+(n/n_c)^{\gamma}\right]^{(\alpha-\beta)/\gamma},
\label{eq_power}
\end{equation}
which applies across the polar band (PB), coexistence (CS), and microcluster (mC) regimes. This form yields $f(n)\sim n^{-\alpha}$ for $n\ll n_c$ and $f(n)\sim n^{-\beta}$ for $n\gg n_c$. In the PB regime (e.g., $g=2,4$), the large-$n$ exponent satisfies $\beta>2$, and $P(n)$ develops an additional peak at large $n$, corresponding to clusters that contain a finite fraction of the total particles (Fig.~\ref{fig_Cn}(b)\,). In this case, the power-law contribution accounts for only a fraction $\sum_{n=1}^{N} n f(n)=\mathcal{N}<N$ of the particles, while the remaining fraction, $1-\mathcal{N}/N$, is contained in a distinct ``infinite-cluster'' peak~\cite{Krishnamurthy1998}.

In contrast, in the mC regime [Fig.~\ref{fig_Cn}(c)], the distribution is fully described by the power-law form $f(n)$ without any separate peak. The asymptotic exponent approaches $\beta\simeq 3/2$ with increasing $g$, and all particles are accounted for within the scale-free distribution $f(n)$.

\subsection{Structural analysis and characterization of the cross-sea phase}\label{sec_VL}
{  
For smaller system sizes, the structures associated with the polar-banded phase disintegrate directly into micro-clusters at higher alignment rates. In contrast, larger systems, which can sustain long-wavelength collective modes, exhibit the emergence of the cross-sea (CS) phase at sufficiently strong alignment; representative morphologies are shown in Fig.~\ref{fig_PD_config}. 
To quantitatively characterize the structural phases, including the CS phase, we compute the Fourier transform of the particle density~\cite{kursten2020dry},
 \begin{equation}
\hat{f}_{\mathbf{k}} := \frac{1}{N} \sum_{i=1}^{N} \exp\left(i \mathbf{k} \cdot \mathbf{r}_i \right)
 \end{equation}
where $\mathbf{k} = \frac{2\pi}{L}(n_x,n_y)$ and $n_x,n_y$ are positive integers. The dominant wave vectors, $\mathbf{k}_1$ and $\mathbf{k}2$, are identified from the largest Fourier amplitudes, $\la |\hat{f}_{\mathbf{k}}| \ra$, and characterize the underlying spatial ordering. Using these modes, we define a lattice order parameter $\ell$ via
\begin{equation}
\ell^2 = \la |\hat{f}_{\mathbf{k}1}| \ra \, \la |\hat{f}_{\mathbf{k}_2}| \ra,
\end{equation}
which measures the simultaneous strength of the two dominant density modulations.
In the banded phase, the density field is typically governed by a single dominant mode, with higher harmonics satisfying relations such as $\mathbf{k}_2 \approx 2\mathbf{k}_1$. Consequently, an order parameter constructed from the two largest Fourier amplitudes alone can attain values comparable to those in the CS phase. To distinguish these phases more reliably, we redefine $\mathbf{k}_2$ by excluding all integer multiples of $\mathbf{k}_1$. Thus, $\mathbf{k}_1$ corresponds to the wave vector with the largest amplitude, while $\mathbf{k}_2$ is chosen as the strongest non-harmonic mode. The emergence of such non-harmonic secondary modes serves as a distinctive signature of the CS phase.

Figure~\ref{fig_fft}(a) presents the time-averaged Fourier amplitudes, $\la |\hat{f}_{\mathbf{k}}| \ra$, in the steady state for different system sizes and parameter values. Figure~\ref{fig_fft}$(a,i)$ corresponds to the PB phase at $\Pe = 10$, $g = 4$, and system size $N = 16000$. The remaining panels, Fig.~\ref{fig_fft}$(a,ii\text{--}vi)$, show the CS phase for $\Pe = 7$ and $g = 5$, with the system size increasing from $N = 16000$ to $N = 256000$. These panels illustrate the gradual stabilization of the CS phase as the system size grows. In particular, the Fourier spectra reveal that larger systems exhibit increasingly regular and well-defined structures, indicating enhanced stability and ordering of the cross-sea patterns.

Figure~\ref{fig_fft}$(b)$ shows the lattice order parameter $\ell$ as a function of system size for $\Pe = 7$ and $g = 5$. For smaller systems, such as $N = 16000$, the order parameter remains relatively low ($\ell \approx 0.22$), indicating that the cross-sea (CS) structure is weakly developed and relatively less stable. With increasing system size, $\ell$ steadily increases and eventually saturates around $\ell \approx 0.36$ near $N=128,000$, signaling the emergence of a stable and well-ordered CS phase.

This pronounced system-size dependence suggests that the formation of extended CS patterns may rely on long-wavelength modes that are suppressed in small simulation boxes, although confirming this mechanism would require a full two-dimensional analysis within a Toner--Tu-like theoretical framework.
The restricted spatial extent suppresses long-range ordering, leading to distorted or fluctuating structures at low $N$. As the system size increases, larger wavelength modes become accessible, enabling the development and stabilization of the CS phase. The corresponding increase in $l$ quantitatively captures this enhanced ordering.

\begin{figure*}[!t]
\centering
\includegraphics[width=17cm]{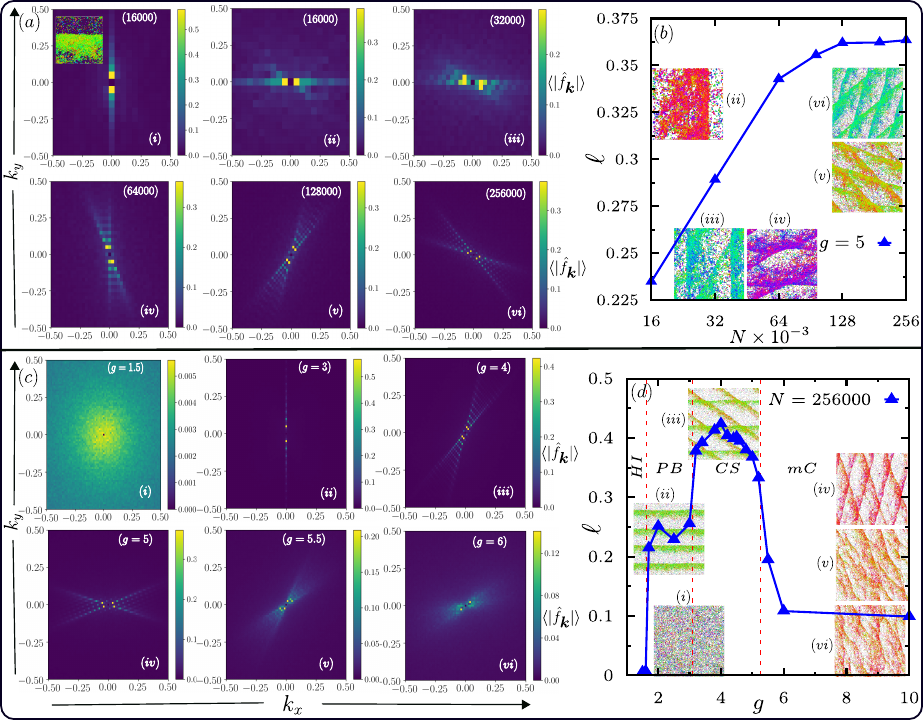}
\caption{
System-size dependence and variation with alignment rate. (a) Absolute value of the Fourier amplitude for (i) a stable banded phase at $(\Pe,g)=(10,4)$ with $N=1.6\times10^4$, and (ii -- vi) the cross-sea phase at $(\Pe,g)=(7,5)$ for $N=1.6\times10^4$, $3.2\times10^4$, $6.4\times10^4$, $1.28\times10^5$, and $2.56\times10^5$, respectively. (b) Lattice order parameter, $\ell$, as a function of system size for $(\Pe,g)=(7,5)$. (c,d) Dependence of the Fourier amplitude and lattice order parameter, respectively, on alignment rate $g$ at $\Pe=7$ for $N=2.56\times10^5$. Red dotted lines in (d) indicate guides to the eye separating the homogeneous isotropic (HI), polar banded (PB), cross-sea (CS), and micro-cluster (mC) regimes. Insets show representative real-space configurations for each phase.
}
\label{fig_fft}
\end{figure*}

Figures~\ref{fig_fft}$(c,d)$ show the time-averaged Fourier spectrum, $\la |\hat{f}_{\mathbf{k}}| \ra$, and the lattice order parameter $\ell$, respectively, as functions of $g$ for $\Pe = 7$ and system size $N = 256000$. 

In the weak-alignment regime, the Fourier spectrum exhibits a broad, diffuse intensity distribution centered around $\mathbf{k}=0$, with no distinct directional peaks. The nearly isotropic spectral distribution corresponds to $\ell \approx 0$, indicating the absence of positional order corresponding to the homogeneous isotropic (HI) phase.

As the alignment rate increases, sharp peaks develop along a single preferred direction in Fourier space, signaling the onset of the banded (PB) phase. At intermediate values of $g$, these dominant peaks split into two distinct directions, marking the emergence of the cross-sea (CS) phase. This phase is characterized by intersecting density bands, enhanced lattice-like ordering, and a pronounced increase in the order parameter $\ell$.

For faster alignment rates, the ordered CS structure becomes unstable and gives way to the micro-cluster (mC) phase. In this regime, the Fourier peaks broaden substantially, while $\ell$ decreases significantly, indicating the breakdown of positional order.

\section{Discussion and Outlook}
\label{sec:conclusion}
We introduced a Vicsek-like Langevin model in which the heading direction of each active particle relaxes toward the local mean orientation over a finite timescale set by the alignment rate $J$, while simultaneously experiencing orientational noise of strength $D_r$. In this sense, the model combines Vicsek-type local consensus with XY-like orientational dynamics. Through extensive numerical simulations, we determined the full phase diagram and demonstrated that the onset of orientational order occurs via a first-order transition. Accompanying this transition, the system exhibits a rich sequence of emergent structural states, including homogeneous isotropic and homogeneous polar phases, as well as polar bands, cross-sea patterns, and micro-clustered states.

The first-order character of the ordering transition is supported by both the Binder cumulant and the pronounced bimodality of the local polarization and density distributions. Near the onset of collective orientational order, the system forms polar bands whose density increases with alignment rate. Their width grows monotonically with $\Pe$ following a power-law dependence $\Pe^{1/2}$ at large activity, while exhibiting a non-monotonic variation with $g$. Upon further increase in the alignment rate, and for sufficiently large system sizes, the system enters a cross-sea phase characterized by a network of intersecting high-density bands. In this regime, the widths of the intersecting bands continue to grow until they eventually merge into a nearly homogeneous polar state. Correspondingly, local density fluctuations are large in the polar-band and cross-sea regimes, decrease upon entering the homogeneous polar phase, and then increase again at still larger alignment rates, signaling the emergence of a micro-clustered state. These morphological transformations were analyzed, with the cross-sea phase in particular characterized quantitatively using the structure factor and a lattice-based order parameter.

Our theoretical study identifies the dimensionless ratio of the alignment timescale, $\tau_{\rm align}$, set by the inverse of the alignment coupling, to the rotational diffusion timescale as a key control parameter governing collective behavior. Direct experimental measurements of $\tau_{\rm align}$, however, remain scarce. Reported estimates range from rapidly aligning starling flocks with $\tau_{\rm align}\approx 0.1\,{\rm s}$~\cite{Mora2016, Nagy2010} to bacterial suspensions exhibiting orientational correlation times of $\sim0.5$--$1.5\,{\rm s}$ and an approximate $1/{\rm Pe}$ scaling \cite{Sokolov_Aranson_2012}, consistent with observations in self-aligning active granular matter \cite{Baconnier2022, Baconnier2025}. Notably, for E. coli, foundational experiments by Berg and Brown established a run time of order $1$ -- $10\,{\rm s}$ together with an effective reorientation timescale of $\sim0.1\,{\rm s}$, providing one of the earliest quantitative characterizations of bacterial orientational dynamics~\cite{Berg_Brown_1972}. Fish schools exhibit orientation-information transfer on timescales of order $1\,{\rm s}$ \cite{Katz2011}, while Quincke rollers display rotational diffusion times of $\sim0.3\,{\rm s}$ together with flocking transitions driven by effective polar alignment interactions \cite{Bricard2013}. Although these measurements remain limited and system dependent, they collectively suggest that increasing activity and alignment rate generally accelerates orientational relaxation, underscoring the need for systematic experimental characterization of both $\tau_{\rm align}$ and rotational diffusivity in order to connect theoretical control parameters with real flocking systems.

Among the morphologies identified here, the polar-band (PB) and homogeneous polar (HP) phases are well-established collective states that emerge within Toner--Tu hydrodynamics~\cite{Solon2015a, caussin2014emergent} and have also been observed experimentally in Quincke-roller systems~\cite{Bricard2013}. In contrast, the cross-sea (CS) phase, with its intrinsically two-dimensional structure, calls for a more general theoretical treatment beyond effectively one-dimensional descriptions. Developing transport-theory and Toner-Tu-type hydrodynamic frameworks capable of capturing the full sequence of states, from homogeneous isotropic (HI) to polar bands (PB) to cross-sea (CS), and subsequently to homogeneous polar (HP) and micro-clustered (mC) phases, therefore, remains an important direction for future work. More broadly, the rich sequence of collective patterns highlights the ability of even a single-species active system to self-organize into a diverse range of nonequilibrium structures. We hope that these results will motivate further theoretical and experimental efforts toward a unified description of these emergent phases and stimulate broader investigations into pattern formation in active matter with finite-time relaxation.

\section*{Data availability}
All data generated or analyzed during this study are included within the article.

\section*{Author Contributions}
D.C. conceived and supervised the study. R.J. and S.S.M. performed the numerical calculations and analyzed the data. D.C. finalized the manuscript with input from R.J. and S.S.M.

\section*{Conflicts of interest}  
There are no conflicts of interest to declare.

\section*{Acknowledgements}
D.C. acknowledges financial support from the Department of Atomic Energy (DAE) through Grant No. 1603/2/2020/IoP/R\&D-II/15028, a Visiting Professorship at CY Cergy Paris Universit{\'e}, and an Associateship of IIT Bombay. D.C. thanks Sriram Ramaswamy, Mustansir Barma, and Fernando Peruani for valuable discussions.
Numerical simulations were performed using the SAMKHYA high-performance computing cluster and other computational facilities at the Institute of Physics, Bhubaneswar.

\appendix
\section{Connection to the Vicsek update rule}
\label{app:vicsek_limit}

Starting from the stochastic angular dynamics
\begin{equation}
d\theta_i = -J\sin(\theta_i-\psi_i)\,dt + \sqrt{2D_r}\,dB_i,
\label{eq:sde_appendix}
\end{equation}
we derive the corresponding Vicsek-type update rule in the strong-alignment limit. Here, $J$ is the alignment rate, $\psi_i$ the local mean orientation, and $D_r$ the rotational diffusivity.

Defining the angular deviation $\delta_i=\theta_i-\psi_i$ and treating $\psi_i$ as constant over an infinitesimal interval gives
$
\frac{d\delta_i}{dt} = -J\sin\delta_i .
$
For strong alignment, $|\delta_i|\ll1$, so that $\sin\delta_i\simeq\delta_i$, yielding
$
\frac{d\delta_i}{dt} =-J\delta_i ,
$
with solution
$
\delta_i(t+dt) = \delta_i(t)e^{-Jdt}.
$
Substituting into Eq.~\eqref{eq:sde_appendix} gives
\begin{equation}
\theta_i(t+dt) = \psi_i(t) + \big[\theta_i(t)-\psi_i(t)\big]e^{-Jdt} + \sqrt{2D_r}\,dB_i .
\label{eq:discrete_appendix}
\end{equation}

In the strong-alignment limit $Jdt\gg1$, the exponential term vanishes and Eq.~\eqref{eq:discrete_appendix} reduces to
\begin{equation}
\theta_i(t+dt) \simeq \psi_i(t) + \sqrt{2D_r}\,dB_i .
\end{equation}
The dynamics, therefore corresponds to instantaneous alignment with the local mean direction followed by angular noise, equivalent to the Vicsek update rule
\begin{equation}
\theta_i(t+dt) = \psi_i(t)+\xi_i ,
\end{equation}
up to the choice of noise distribution.

\section{Role of alignment rate and rotational diffusion constant}\label{app_alignment}

\begin{figure}[!h]
\centering
 \includegraphics[width=4cm]{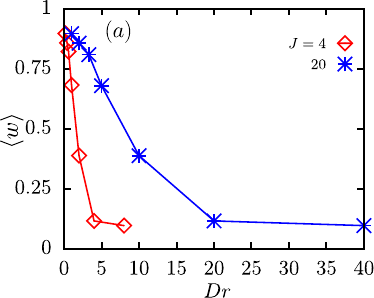}
 \includegraphics[width=4cm]{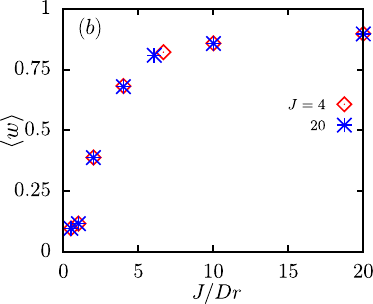}
 \caption{(a)~Average polar order $\langle w \rangle$ for two values of $J$ as a function of rotational diffusion $D_r$ at $\mathrm{Pe} = 10$. (b)~The data collapse onto a single curve when plotted versus the rescaled coupling $g = J/D_r$, for system size $N = 4000$ and particle density $\rho = 1.0$.
 }\label{fig_ScaledGr}
\end{figure}

In the main text, we assumed that the alignment strength $J$ and the rotational diffusivity $D_r$ do not independently control the system's properties, as in equilibrium, where only appropriate dimensionless combinations are relevant. Here, we test this assumption.
We compute the polar order parameter $\langle w \rangle$ as a function of $D_r$ for two values of $J$ [Fig.~\ref{fig_ScaledGr}(a)]. The resulting curves show distinct dependences on $D_r$. However, as shown in Fig.~\ref{fig_ScaledGr}(b), the data collapse onto a single curve when plotted versus the rescaled parameter $g = J/D_r$. This collapse demonstrates that $g$ is the relevant control parameter governing orientational order and, consequently, the macroscopic behavior of the system.

\bibliographystyle{unsrt}

\end{document}